# Exact Black-Hole Solution With Self-Interacting Scalar Field

Olaf Bechmann and Olaf Lechtenfeld

*Institut für Theoretische Physik, Universität Hannover*
*Appelstraße 2, 30167 Hannover, Germany*
E-mail: bechmann, lechtenf@itp.uni-hannover.de

## Abstract

Einstein gravity minimally coupled to a self-interacting scalar field is investigated in the static and isotropic situation. We explicitly construct in partially closed form a new black-hole solution with exponentially decaying scalar hair. The symmetric interaction potential has both signs and a triple-well shape with a smooth but non-analytic minimum at vanishing field. We present numerical data as well as double series expansions around spatial infinity.

It is known for more than 20 years that isotropic and static solutions of Einstein's equations are very rigid in nature. In vacuo, with $T_{\mu\nu}=0$, where isotropy already implies time-independence, the Schwarzschild metric $g^{(s)}_{\mu\nu}(r)$ is in fact the *unique* asymptotically flat solution, depending on the two parameters $r_0$ (location of the singularity) and $r_s$ (location of the event horizon). The situation is less clear in the presence of matter, although partial results exist for gravity coupled to Maxwell, Yang-Mills, and/or scalar fields of dilaton, axion or Higgs type. For a review see ref. [1]. The so-called "no-hair" theorems severely restrict the static field configurations outside the horizon, completely classifying regular and asymptotically flat black-hole solutions by a few conserved charges such as total mass, angular momentum, electric and magnetic charges [2].

The most simple example is that of a minimally gravitationally coupled real scalar field $\phi$ enjoying some self-interaction $V(\phi)$. We are interested in spherically symmetric and static field configurations $(g_{\mu\nu}(r), \phi(r))$, where $r$ denotes the radial coordinate. The Minkowksi metric $ds^2 = -dt^2+dr^2+r^2d\Omega^2$ determines our sign convention. As $r \to \infty$, the scalar function $\phi(r)$ approaches a constant which we may normalize to zero. Asymptotic flatness then demands that $V(0) = V'(0) = 0$, i.e. a vanishing cosmological constant and a local extremum of the interaction potential at the origin $\phi$=0.

For this case a scalar no-hair theorem can be demonstrated by a simple argument due to Bekenstein [3]. The scalar field equation, [1]

$$\partial_\mu \sqrt{g} g^{\mu\nu} \partial_\nu \phi \;=\; \sqrt{g}\, V'(\phi) \quad , \tag{1}$$

receives only contributions from $\mu = \nu = r$. After multiplying with $\phi(r)$ and integrating from the horizon to infinity, a partial integration yields

$$[\phi\, g^{rr} \sqrt{g} \partial_r \phi]\Big|_h^\infty - \int_h^\infty dr\, \sqrt{g} g^{rr} (\partial_r \phi)^2 \;=\; \int_h^\infty dr\, \sqrt{g}\, \phi\, V'(\phi) \quad . \tag{2}$$

The horizon $h$ is defined by the largest zero of $g^{rr}$, so that $g^{rr} \geq 0$ in the integration domain. Assuming regularity of $\sqrt{g}$ and $\phi$ at the horizon as well as a fall-off of $\phi(r) = o(r^{-1/2})$ for $r \to \infty$, we can drop the boundary terms in eq. (2). The integrands on both sides of the equation are clearly non-negative, provided that

$$\phi(r)\, V'(\phi(r)) \;\geq\; 0 \qquad \text{for} \quad r \geq h \quad . \tag{3}$$

Since the left-hand integral of eq. (2) comes with a negative sign, both integrals must be zero and, hence, both integrands vanish identically. It follows that the scalar field sits at its asymptotic value, $\phi(r) \equiv 0$, so that merely the Schwarzschild metric results. This eliminates the possibility of non-trivial scalar deformations of the Schwarzschild black hole. Besides the reasonable regularity and fall-off assumptions above, the non-trivial condition going into this no-hair argument is eq. (3) which means that the potential function has a local minimum at $\phi$=0 and must not have a *local maximum* in the $\phi$-range probed outside the horizon.

[1] We abbreviate $\sqrt{g} = \sqrt{-\det g_{\rho\lambda}}$.

Indeed, eq. (3) can be weakend. [2] Heusler [4] has shown that the *dominant energy condition* is in conflict with spherically symmetric scalar field perturbations of the Schwarzschild solution. In our context the dominant energy condition reads $V(\phi) \geq 0$, so that only a *non-negative potential* is required to rule out scalar hair.

In this paper we shall look at the situation where the positivity condition and thus eq. (3) is violated. Is it possible to improve on the scalar no-hair theorem by relaxing or dropping this condition? Or can one find a non-trivial solution for some indefinite potential, showing that eq. (3) is indeed essential? We shall provide conclusive evidence for the second choice by explicitly constructing such a solution of the coupled Einstein-scalar equations.

Our starting point is the generalized Einstein-Hilbert action

$$S[g,\phi] \;=\; \int d^4x\,\sqrt{g}\,[R+\tfrac{1}{2}g^{\mu\nu}\partial_\mu\phi\partial_\nu\phi+V(\phi)] \quad . \tag{4}$$

It is extremized by eq. (1) and

$$R_{\mu\nu}-\tfrac{1}{2}g_{\mu\nu}R \;=\; -\tfrac{1}{2}\partial_\mu\phi\partial_\nu\phi+\tfrac{1}{2}g_{\mu\nu}(\tfrac{1}{2}\partial\phi\cdot\partial\phi+V) \;\equiv\; -T_{\mu\nu} \tag{5}$$

which may be simplified to

$$R_{\mu\nu}+\tfrac{1}{2}\partial_\mu\phi\partial_\nu\phi+\tfrac{1}{2}g_{\mu\nu}V(\phi) \;=\; 0 \quad . \tag{6}$$

In the isotropic and static case all field degrees of freedom are functions of the radial coordinate $r$ only, and the metric can be reduced to two functions by residual coordinate transformations. The field configuration is then given by

$$\phi \;=\; \phi(r) \qquad \text{and} \qquad ds^2 \;=\; -G(r)\,dt^2+G(r)^{-1}dr^2+S(r)^2d\Omega^2 \tag{7}$$

where we have chosen a somewhat non-standard gauge [5]. We often use

$$\sigma(r) \;=\; -\ln S(r) \qquad\qquad \Leftrightarrow \qquad\qquad S(r) \;=\; e^{-\sigma(r)} \tag{8}$$

instead of $S(r)$. Equations (1) and (6) reduce to four coupled ordinary second-order differential equations, three of which are independent. We chose

$$\begin{aligned} \sigma'' \;-\; \sigma'^2 \;+\; \tfrac{1}{4}\phi'^2 \;&=\; 0 \\ G'' \;-\; 2G'\sigma' \;-\; V(\phi) \;&=\; 0 \\ G'' \;-\; 2G\,(2\sigma'^2-\sigma'') \;+\; 2e^{2\sigma} \;&=\; 0 \quad , \end{aligned} \tag{9}$$

[2] We thank M. Heusler for pointing this out to us.

where the prime denotes a derivative with respect to $r$. The scalar's equation of motion follows from these. We take the potential to be an *even* function, so that the sign of the scalar field is undetermined. Given the potential $V(\phi)$, eqs. (9) determine the three functions $\phi(r)$, $G(r)$ and $\sigma(r)$ (or $S(r)$). We prescribe their asymptotical behavior for $r \to \infty$ as

$$\phi(r) \;\to\; 0 \qquad , \qquad G(r) \;\sim\; 1 - \frac{2M}{r} \qquad , \qquad \sigma(r) \;\sim\; -\ln r \quad , \tag{10}$$

with the black-hole mass $M > 0$ being an integration constant. Finally, we fix the freedom of shifting $r$, either by taking $S(0) = 0$, or by putting $S(r) - r = O(\frac{1}{r})$ for $r \to \infty$.

The solution to eqs. (9) is known analytically only when the scalar field is free and massless. Putting $V(\phi) \equiv 0$, we can solve the second and third of eqs. (9) for $G(r)$ in terms of $\sigma(r)$. Eliminating $G(r)$ one arrives at an equation for $\sigma(r)$ only,

$$(r - \alpha h)\,\sigma''(r) \;-\; (2r - h)\,\sigma'^2(r) \;-\; \sigma'(r) \;=\; 0 \quad , \tag{11}$$

with integration constants $\alpha$ and $h$ satisfying $\alpha \geq 1$. This homogeneous Ricatti equation for $\sigma'(r)$ is easily solved by

$$\begin{aligned} \sigma(r) &= -\alpha \ln r - (1-\alpha)\ln(r-h) \\ &= -\ln r - (1-\alpha)\ln(1 - \tfrac{h}{r}) \quad , \end{aligned} \tag{12}$$

where two further integration constants are fixed by the $r{\to}0$ and $r{\to}\infty$ boundary conditions and taking the left-most singularity to sit at $r{=}0$. Equations (9) and (10) then directly yield [3]

$$\begin{aligned} G(r) &= (1 - \tfrac{h}{r})^{2\alpha-1} \\ \phi(r) &= \pm 2\sqrt{\alpha(\alpha{-}1)}\,\ln(1 - \tfrac{h}{r}) \quad , \end{aligned} \tag{13}$$

so that $M = (\alpha{-}\frac{1}{2})h$. This two-parameter family of solutions was found by Buchdahl [6] already in 1959, though by a completely different route. At $r = h$, however, the scalar field develops a physical singularity, so that the no-hair theorem above is avoided. The only exception obtains for $\alpha = 1$ and is, as expected, nothing but the Schwarzschild metric with horizon at $r = h$ and $\phi(r) \equiv 0$.

When the potential $V(\phi)$ is not constant, an analytic solution to eqs. (9) seems hard to come by, even for simple cases of $V$. To keep the choice of potential open, it is useful to translate the $V$-dependence in eqs. (9) to a fourth function

$$U(r) \;:=\; V(\phi(r)) \tag{14}$$

[3] The location $h$ of the second singularity is then positive.

so that a solution is given by the quartet $(\phi, G, \sigma, U)$. Now it turns out to be fruitful to reverse the roles of the potential and the scalar field in our problem. In other words, we are going to first choose some fixed scalar field configuration $\phi(r)$ and then seek to determine the corresponding metric and potential function $U(r)$ from which the field potential $V(\phi)$ can be reconstructed. [4] The advantage of this approach is that the first of eqs. (9) is the simplest and can actually be solved analytically for a certain function $\phi(r)$.

To be more precise, the inhomogeneous Ricatti equation

$$\sigma''(r) \; - \; \sigma'(r)^2 \; + \; \tfrac{1}{4}\phi'(r)^2 \; = \; 0 \tag{15}$$

is analytically soluble for the ansatz

$$\phi(r) \; = \; \phi_0 \, e^{-mr} \qquad\qquad \text{with} \quad m > 0 \quad . \tag{16}$$

One finds [5]

$$m\,S(r) \; \equiv \; m\,e^{-\sigma(r)} \; = \; K_0(\tfrac{1}{2}\phi(r)) \; + \; (\ln\tfrac{\phi_0}{4} + \gamma)\,I_0(\tfrac{1}{2}\phi(r)) \tag{17}$$

where $K_0$ and $I_0$ are the modified Bessel functions, and $\gamma$=0.577... is the Euler-Mascheroni constant. Two integration constants are fixed by demanding that $S(r{\to}\infty) \sim r + O(\phi^2)$. The location $r_0$ of the physical singularity is then given by the largest zero of $S$, i.e. $S(r_0) = 0$. [6]

The associated functions $G(r)$ and $U(r)$ can in principle be computed by going into the remaining two of eqs. (9), which may be brought to the integral form

$$\begin{aligned} G(r) \; &= \; S(r)^2 \int_r^\infty dr' \, \frac{2r' - 6M}{S(r')^4} \\ U(r) \; &= \; \tfrac{1}{2} G(r) \, \frac{(S(r)^4)''}{S(r)^4} \; + \; (2r - 6M) \Big(\frac{1}{S(r)^2}\Big)' \; - \; \frac{2}{S(r)^2} \quad , \end{aligned} \tag{18}$$

with one further integration constant $M$ chosen such that $G(r) \sim 1 - \frac{2M}{r}$ asymptotically. It turns out that $r_0 \leq 0 < M$, so that $G(r)$ develops a single pole at $r{=}r_0$, with negative residue. The black-hole mass $M$ is related to the event horizon $h$ by

$$3M \; = \; \frac{\int_h^\infty dr \; r\,S(r)^{-4}}{\int_h^\infty dr\, S(r)^{-4}} \quad , \tag{19}$$

where the latter is defined by $G(h) = 0$. It follows that $r_0 < h < 3M$ so that the singularity is shielded. It is convenient to set $h = 1$ which amounts to measuring distances

[4] in the region where the chosen $\phi$ takes values.

[5] We take $\phi_0 > 0$ for simplicity.

[6] Of course, we could shift $r_0$ to zero, at the expense of adding the constant $r_0$ to the asymptotic behavior of $S$. Numerically, one finds $r_0 = -0.0686\ldots$ for $\phi_0{=}m{=}1$.

in units of horizon lengths. The remaining parameters of our solution are the scalar field amplitude $\phi_0$ and its mass parameter $m$.

Expanding the Bessel functions for small argument (around $r{=}\infty$), eq. (17) becomes

$$S(r) \;=\; r \;+\; \sum_{k=1}^{\infty} \frac{1}{k!^2}\,(\tfrac{1}{4}\phi_0)^{2k} \left[ r + \frac{1}{m} \sum_{j=1}^{k} \frac{1}{j} \right] e^{-2kmr} \quad . \tag{20}$$

Equation (18) then produces series expansions in the form

$$\sum_{k=1}^{\infty} r\, f_k(\tfrac{1}{r})\, e^{-2kmr} \;+\; \sum_{k=1}^{\infty}\sum_{\ell=1}^{\infty} r^2\, f_{k\ell}(\tfrac{1}{r})\, Ei(-2\ell mr)\, e^{-2kmr} \tag{21}$$

for $G(r) - (1-\frac{2M}{r})$ and $U(r)$, with polynomials $f_k$ and $f_{k\ell}$. We have evaluated the expansions à la eq. (21) up to order $e^{-10mr}$ and tested them in the original eqs. (9). The numerical accuracy outside the horizon is generally of $O(10^{-10})$ and decreases at the horizon to $O(10^{-7})$.

It is instructive to plot the deviation of $S(r)$ and $G(r)$ from the Schwarzschild case. This is done in Figs. 1 and 2, for $h{=}1$, $\phi_0{=}1$, and three different values of $m$. It is evident that the metric is regular for $r{>}r_0$ except for the standard Schwarzschild coordinate singularity at $r{=}1$. [7]

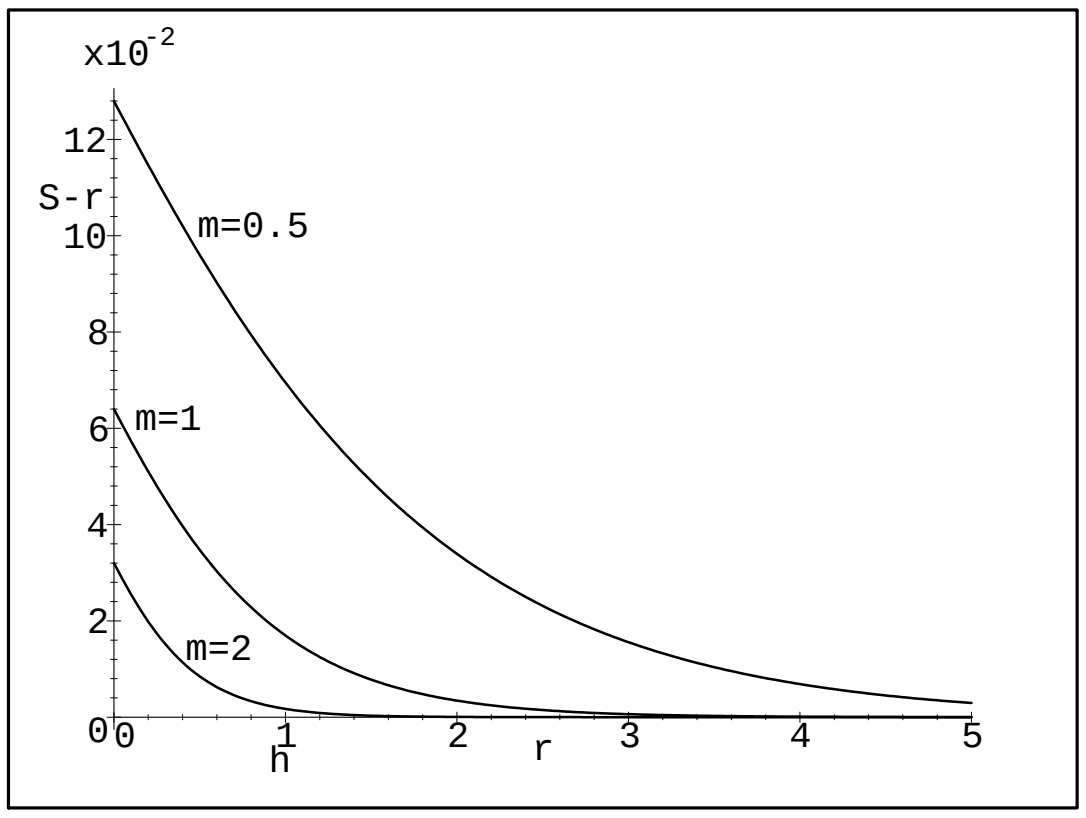

Fig. 1: Deviation of $S(r)$ from Schwarzschild case

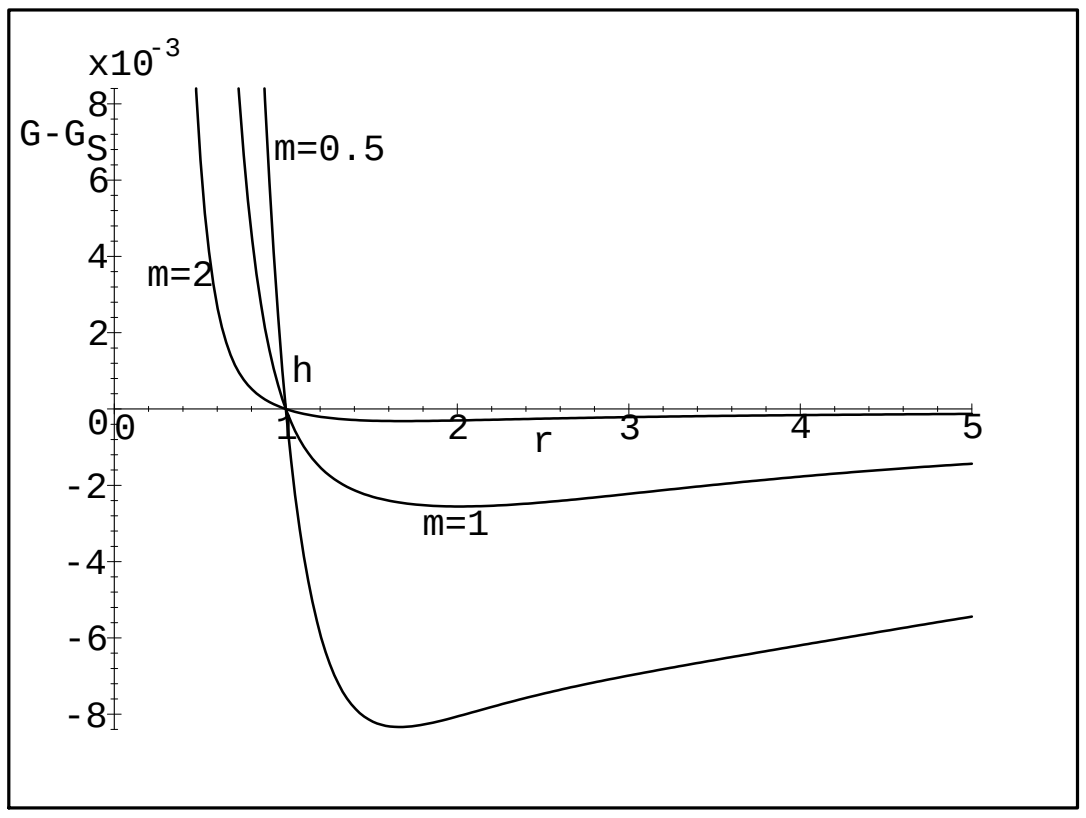

Fig. 2: Deviation of $G(r)$ from Schwarzschild case

The most interesting object is, of course, the potential. Fig. 3 shows the function $U(r)$ for $\phi_0{=}1$ and three values of $m$. A closer look reveals that $U(r)$ has a local *maximum* and for $r \to \infty$ approaches zero *from above*. Note also that $U < 0$ in the vicinity of the horizon.

[7] For large enough amplitudes $\phi_0$ the metric function $G(r)$ develops a new zero which can then be taken as the event horizon, without qualitatively changing the discussion.

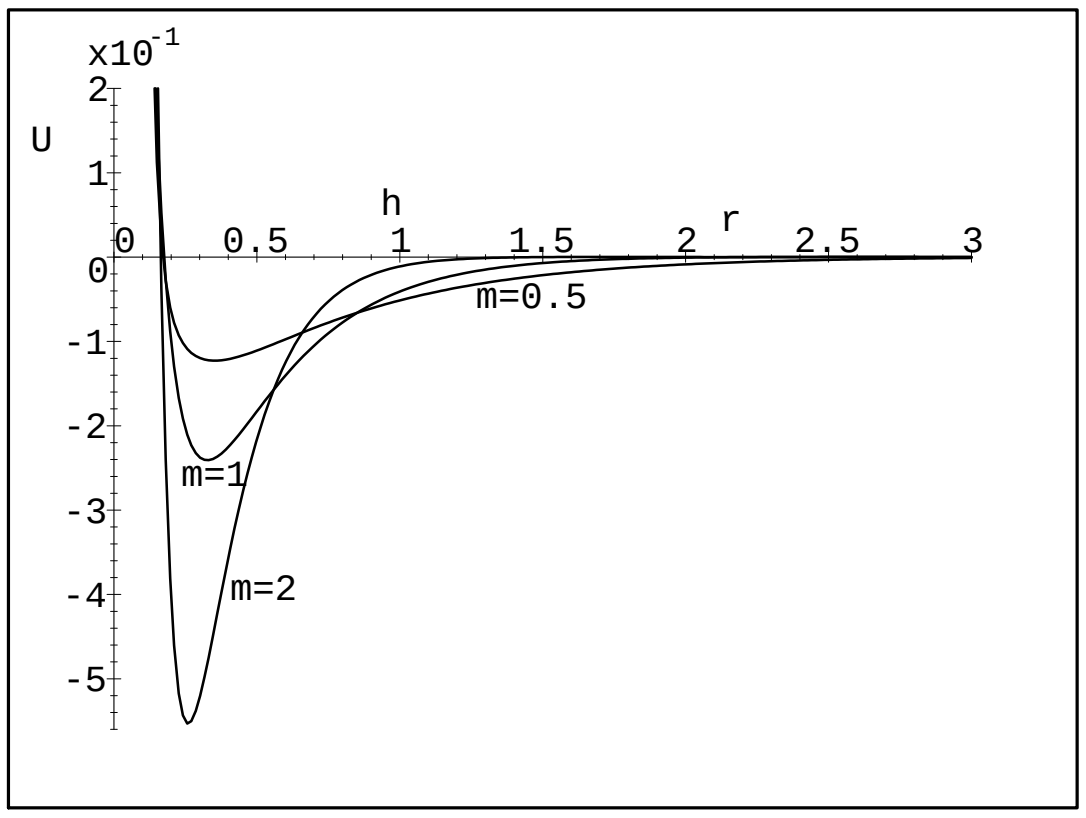

Fig. 3a: Potential function $U(r)$

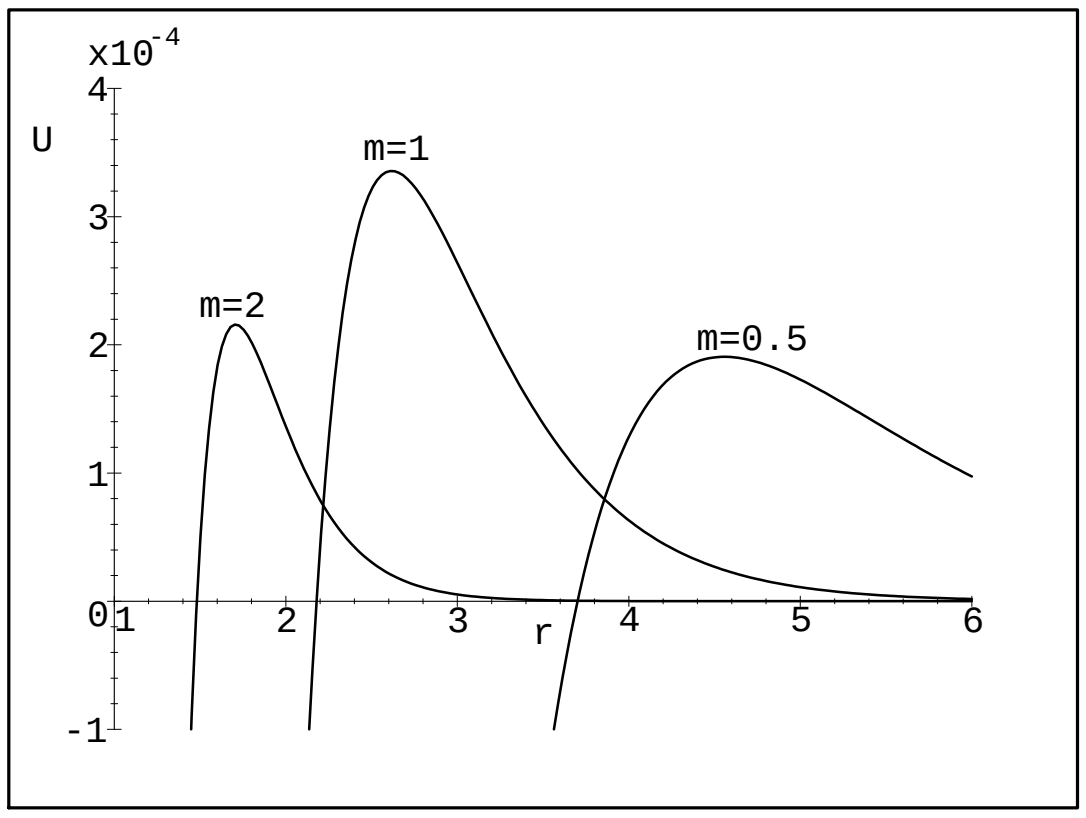

Fig. 3b: Detail of $U(r)$

Composing $U(r)$ with $r(\phi) = -\frac{1}{m}\ln|\phi/\phi_0|$, one arrives at Fig. 4 which displays the interaction potential $V(\phi)$ in the range $\phi \in [0, \phi_0{=}1]$. The interesting region is blown up to show that $V$ has indeed a local *minimum* at the origin. Reflection at $\phi{=}0$ extends the function to negative $\phi$, so that a triple-well potential results. $G(r)$ and $U(r)$ diverge at $r{=}r_0$, so that $V(\phi)$ explodes like $\ln^{-3}|\phi/\phi_1|$ for $\phi \to \phi_1 {\equiv} \phi_0 e^{-mr_0}$. Clearly, our solution escapes the consequences of the no-hair theorem by having a partially negative potential, demonstrating that the dominant energy condition is a necessary one.

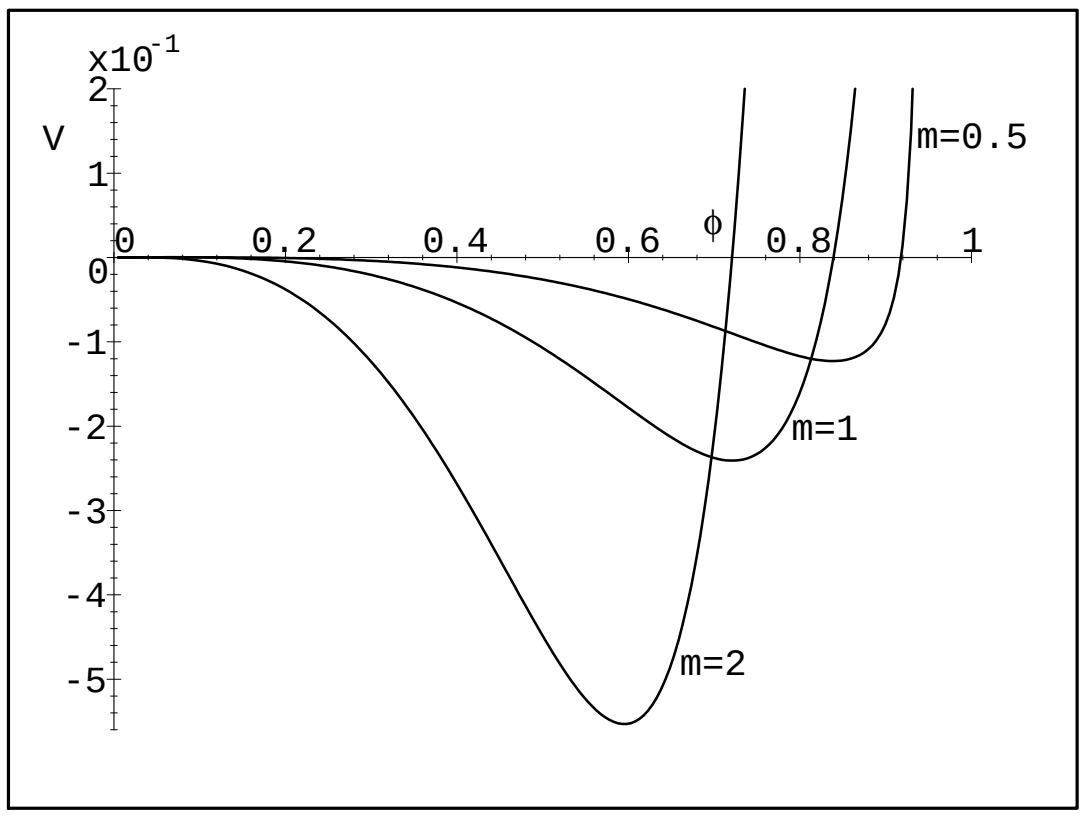

Fig. 4a: Interaction potential $V(\phi)$

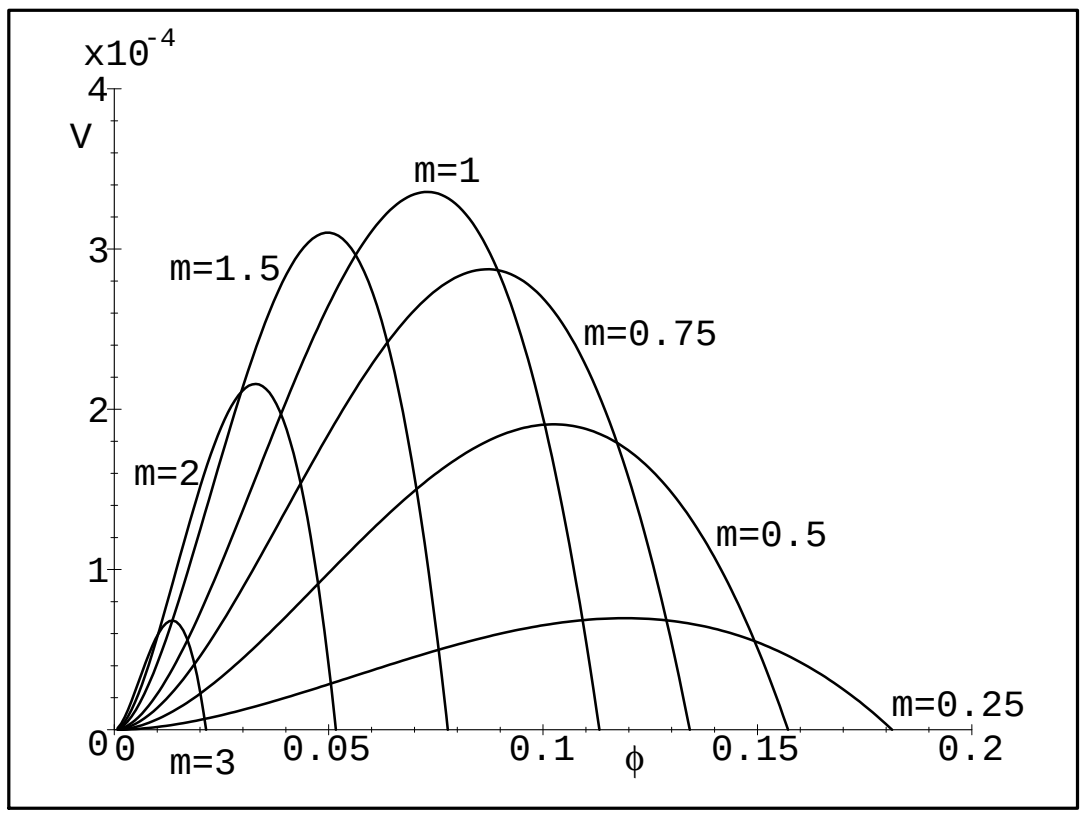

Fig. 4b: Detail of $V(\phi)$

In spherical coordinates, the energy-momentum tensor of our solution reads

$$T_{\mu\nu} \;=\; \mathrm{diag}\,(\rho G\,,\, p_r G^{-1}\,,\, p_t S^2\,,\, p_t S^2 \sin^2\theta) \tag{22}$$

with energy density $\rho$, radial pressure $p_r$ and tangential pressure $p_t$ given by

$$\begin{aligned} \rho &= +\tfrac{1}{4}G\,\phi'^2 + \tfrac{1}{2}V \\ p_r &= +\tfrac{1}{4}G\,\phi'^2 - \tfrac{1}{2}V \\ p_t &= -\tfrac{1}{4}G\,\phi'^2 - \tfrac{1}{2}V \quad . \end{aligned} \tag{23}$$

Two different types of pressure occur since black-hole configurations are isotropic but not homogeneous [7]. Energy density and pressures are plotted for $\phi_0{=}m{=}1$ as functions of $r$ in Fig. 5 and are seen to be regular for $r > r_0$. Near the horizon the radial pressure dominates the energy density since the latter turns negative due to the negative potential there.

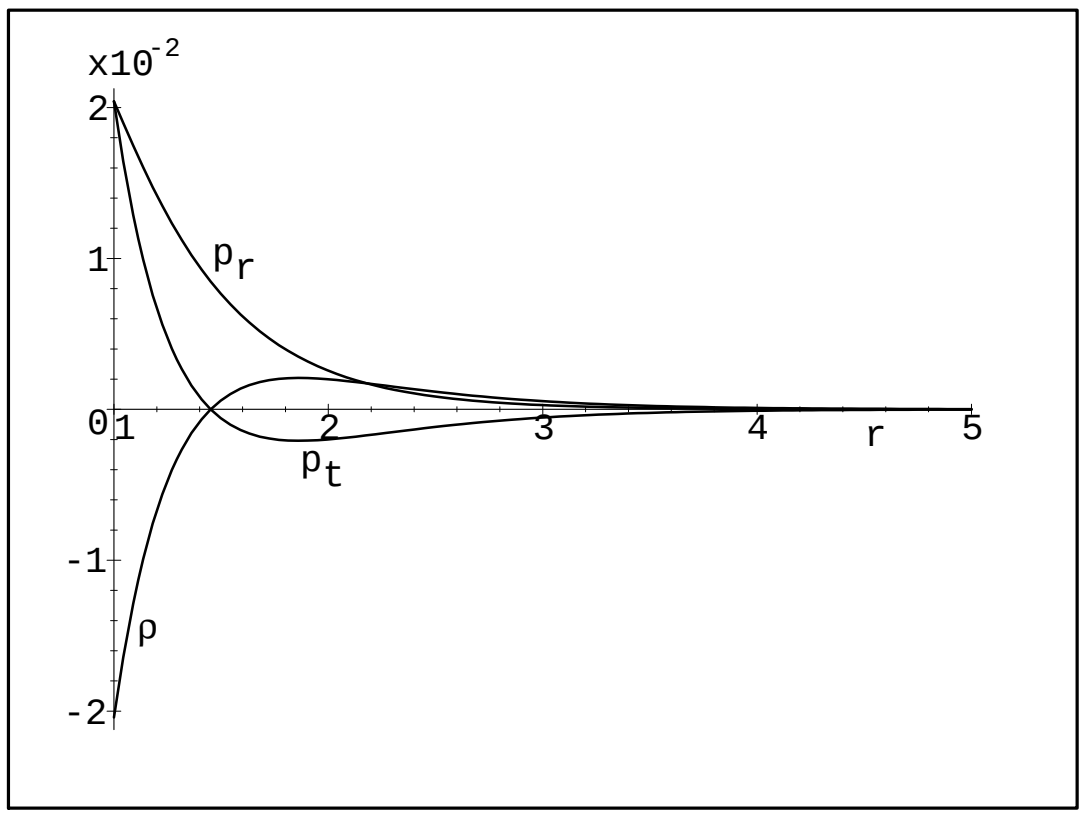

Fig. 5a: Functions $\rho$, $p_r$ and $p_t$ outside the horizon

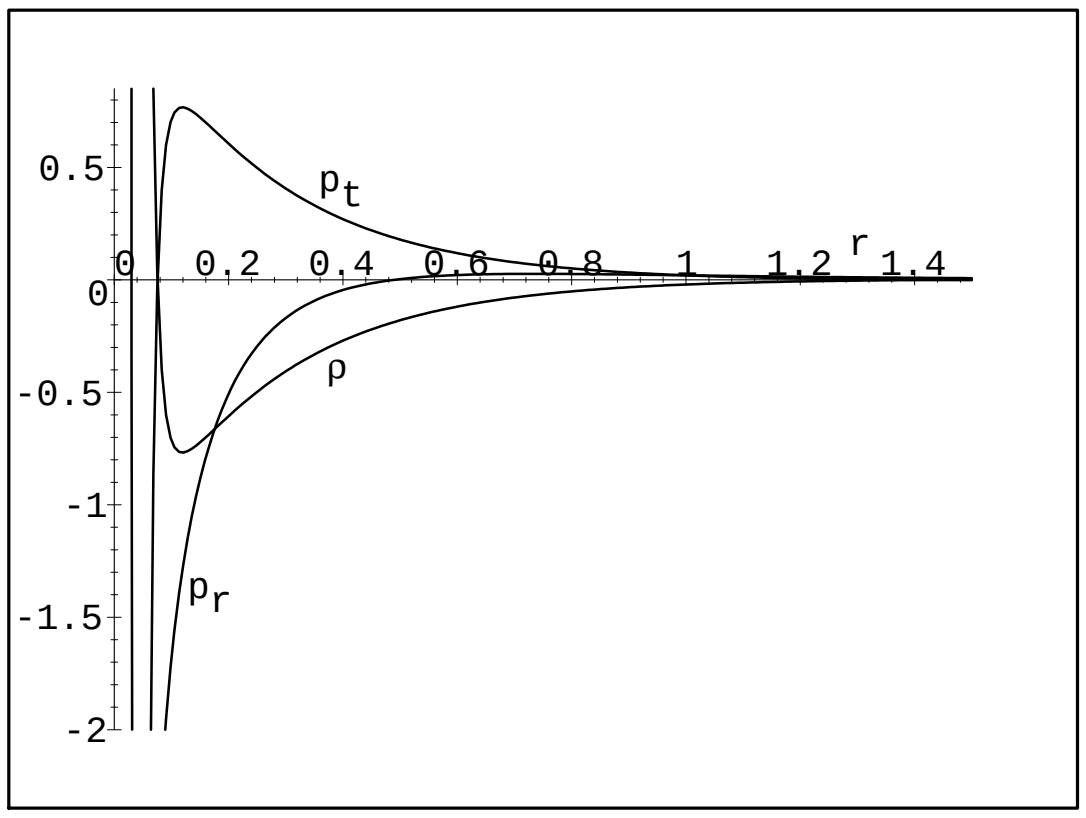

Fig. 5b: Functions $\rho$, $p_r$ and $p_t$ inside the horizon

In order to verify the consistency of our solution, we finally develop an power series expansion around $r = \infty$. Let us first discuss the expansion of a general solution to eqs. (9). The natural expansion parameter is $\frac{1}{r}$. Naively assuming analyticity of all functions in $\frac{1}{r}$ we attempt [8]

$$S(r) \;=\; r \;+\; \sum_{i=1}^{\infty} s_i\, r^{-i} \tag{24}$$

and plug this into

$$\phi'(r)^2 \;=\; 4\,S''(r)/S(r) \tag{25}$$

and eqs. (18) to obtain power series for $\phi$, $G$ and $U$. The leading terms of these expansions come out to be

$$\phi(r) \;=\; O(r^{-k}) \qquad \Longrightarrow \qquad U(r) \;=\; O(r^{-2k-2}) \qquad \Longrightarrow \qquad V(\phi) \;=\; O(|\phi|^{2+2/k}) \tag{26}$$

[8] A constant ($s_0$) term has been set to zero to fix an integration constant, in agreement with eqs. (18).

so that only $k$=1 may lead to a potential $V$ with a local *analytic* minimum at the origin. However, in this case the leading coefficient of $U(r)$ vanishes and, hence, $V(\phi) \sim |\phi|^5$. Clearly, we have to go beyond simple power series and give up analyticity at $r$=$\infty$.

On the other hand, eqs. (16), (20) and (21) show that our solution is essentially "non-perturbative", since $e^{-mr}$ is not analytic in the expansion parameter $\frac{1}{r}$. Hence, a double expansion, in $\frac{1}{r}$ and $e^{-mr}$, is needed. Keeping only the leading non-trivial order in $e^{-mr}$, we write down an improved general ansatz

$$\begin{aligned}
\phi(r) &= e^{-mr}\, r^a \sum_{i\geq 0} f_i\, r^{-i} \;+\; O(e^{-2mr}) \\
\sigma(r) &= -\ln r \;+\; e^{-2mr}\, r^b \sum_{i\geq 0} \sigma_i\, r^{-i} \;+\; O(e^{-4mr}) \\
G(r) &= 1 - \frac{2M}{r} \;+\; e^{-2mr}\, r^c \sum_{i\geq 0} g_i\, r^{-i} \;+\; O(e^{-4mr}) \\
U(r) &= e^{-2mr}\, r^d \sum_{i\geq 0} u_i\, r^{-i} \;+\; O(e^{-4mr})
\end{aligned} \tag{27}$$

and insert into eqs. (9) to obtain equations for the coefficients $f_i$, $\sigma_i$, $g_i$, and $u_i$. The leading powers come out to be

$$2a \;=\; b \;=\; c \;=\; d \quad . \tag{28}$$

Our solution was obtained by taking

$$a \;=\; 0 \qquad\qquad \text{and} \qquad\qquad f_i \;=\; \phi_0\, \delta_{i0} \qquad \forall i \tag{29}$$

which yields

$$\sigma_0 \;=\; -\tfrac{1}{16}\,\phi_0^2 \quad , \qquad \sigma_1 \;=\; -\tfrac{1}{16m}\,\phi_0^2 \quad , \qquad \sigma_i \;=\; 0 \qquad \forall i \geq 2 \tag{30}$$

and

$$\begin{array}{lll}
g_0 = +\tfrac{1}{8}\,\phi_0^2 & \qquad & u_0 = +\tfrac{1}{2}m^2\,\phi_0^2 \\
g_1 = -\tfrac{1}{8}m^{-1}\,(1{+}2Mm)\,\phi_0^2 & & u_1 = -m\,(1{+}Mm)\,\phi_0^2 \\
g_2 = +\tfrac{1}{8}m^{-2}\,(1{+}4Mm)\,\phi_0^2 & & u_2 = +\tfrac{1}{2}\,(1{+}3Mm)\,\phi_0^2 \\
g_3 = -\tfrac{1}{4}m^{-3}\,(1{+}3Mm)\,\phi_0^2 & & u_3 = -\tfrac{1}{2}m^{-1}\,(1{+}3Mm)\,\phi_0^2 \\
\vdots & & \vdots
\end{array} \quad . \tag{31}$$

This result agrees perfectly with the series in eqs. (20) and (21),

$$\begin{aligned}
\sigma(r) &= -\ln r \;-\; \tfrac{1}{16}\,\phi_0^2\, e^{-2mr}\Big[1 + \tfrac{1}{mr}\Big] \;+\; O(e^{-4mr}) \\
G(r) &= 1 - \frac{2M}{r} \;+\; \tfrac{1}{3}\phi_0^2\, r^2\, Ei(-2mr)\, m^2(1{+}3Mm) \\
&\quad + \tfrac{1}{24}\phi_0^2\, e^{-2mr}\Big[4m(1{+}3Mm)\,r + (1{-}6Mm) - \tfrac{1}{m}\tfrac{1}{r} + \tfrac{3M}{m}\tfrac{1}{r^2}\Big] \;+\; O(e^{-4mr}) \\
U(r) &= 2\phi_0^2\, Ei(-2mr)\, m^2(1{+}3Mm) \;+\; \tfrac{1}{2}\phi_0^2\, e^{-2mr} m^2\Big[1 + \tfrac{4M}{r}\Big] \;+\; O(e^{-4mr}) \quad ,
\end{aligned} \tag{32}$$

after expanding the exponential-integrals.

Let us take a look at the potential $V(\phi)$. The non-analyticities of $U(r) \sim e^{-2mr}$ and $r(\phi) = -\frac{1}{m}\ln|\phi/\phi_0|$ tend to compensate each other so that $V(\phi) \sim \frac{1}{2}m\phi^2$. However, the additional power series $\sum_i u_i\, r^{-i}$ introduces logarithmic terms into the potential

$$V(\phi) \;=\; \tfrac{1}{2}m\,\phi^2\left[1+\frac{2+2Mm}{\ln|\phi/\phi_0|}+\frac{1+3Mm}{\ln^2|\phi/\phi_0|}+\ldots\right] \;+\; O(\phi^4) \tag{33}$$

which are necessary to match the expansions. It should be noted that $V(\phi)$ is nevertheless $\mathbf{C}^\infty$ smooth at $\phi{=}0$. This exemplifies how non-trivial solutions require a non-analyticity at $r{=}\infty$.

To summarize, we have constructed an explicit static and isotropic black-hole solution to Einstein's equations minimally coupled to a self-interacting scalar field. It is regular outside the central singularity, with the standard coordinate singularity signalling a regular event horizon. An exponentially decaying scalar field configuration belongs to an interaction potential which, firstly, is sometimes negative and, secondly, is perfectly smooth but not analytic at an (asymptotically attained) local minimum. The first feature circumvents the no-hair theorem while the second one seems to be generic. It would be interesting and physically relevant to investigate the stability properties of this solution. Work in this direction is in progress.

Acknowledgment:

O.L. is grateful to N. Dragon, F. Müller-Hoissen, S. Samuel and E. Weinberg for discussions. We thank M. Heusler and C. Lousto for drawing our attention to refs. [4] and the third of ref. [6].

# References

[1] G.W. Gibbons, *Self-gravitating magnetic monopoles, global monopoles and black holes*, 12th Lisbon Autumn School on Physics, Springer-Verlag, 1991.

[2] V. Ginzburg and L. Ozernoi, Soviet Phys. JETP **20** (1965) 689;
A. Doroshkevich, Ya. Zeldovich and I. Novikov, Sov. Phys. JETP **22** (1966) 122;
W. Israel, Phys. Rev. **164** (1967) 1776, Commun. Math. Phys. **8** (1968) 245;
B. Carter, Phys. Rev. Lett. **26** (1971) 331;
J.D. Bekenstein, Phys. Rev. **D5** (1972) 1239, *ibid.* 2403;
J.B. Hartle, in *Magic without Magic*, ed. J. Klauder (Freeman, 1972);
C. Teitelboim, Phys. Rev. **D5** (1972) 2941;
R.H. Price, Phys. Rev. **D5** (1972) 2419, *ibid.* 2439.

[3] J.D. Bekenstein, in [2].

[4] M. Heusler, J. Math. Phys. **33** (1992) 3497, Class. and Qu. Grav. **10** (1993) 791, U. Chicago preprint (Nov. 94), gr-qc/9411054.

[5] D. Garfinkle, G.T. Horowitz and A. Strominger, Phys. Rev. **D43** (1991) 3140.

[6] H. Buchdahl, Phys. Rev. **115** (1959) 1325;
A.I. Janis, D.C. Robinson and J. Winicour, Phys. Rev. **186** (1969) 1729;
M. Wyman, Phys. Rev. **D24** (1981) 839;
O.A. Fonarev, Hebrew University preprint (Sept. 1994), gr-qc/9409020.

[7] F.V. Kusmartsev, E.W. Mielke, F.E. Schunck, Phys. Rev. **D42** (1990) 3388.